\documentclass[]{spie}  
\usepackage[]{graphicx}
\usepackage[utf8]{inputenc}
\usepackage{import}
\usepackage{url}
\usepackage[margin=1cm]{caption}
\usepackage{makecell}

\usepackage{scrlayer}
\DeclareNewLayer{header}
\newcommand*\thispageonlythisheader[1]{%
	\DeclareLayer[background,head,contents={#1}]{header}%
	\DeclarePageStyleByLayers{header}{header}%
	\thispagestyle{header}%
}

\title{ 
	\vspace{.50cm}
	Transfer Learning in Automated Gamma Spectral Identification} 

\author{Eric T. Moore\supit{a} and Johanna Turk\supit{a},\\
	William P. Ford\supit{b}, Nathan J. Hoteling\supit{c}, and Lance S. McLean\supit{d} 
	\skiplinehalf
	\supit{a}Special Technologies Laboratory, Santa Barbara, CA \\
	\supit{b}The Probitas Project, Inc., Vienna, VA \\
	\supit{c}Remote Sensing Laboratory, Joint Base Andrews, MD \\
	\supit{d}Noblis ESI, Chantilly, VA
}

\authorinfo{Further author information: (Send correspondence to E.T. Moore)\\ E.T. Moore: E-mail: mooreet@nv.doe.gov, Telephone: 1 703 853 7994}

\begin{document} 
	\maketitle
	
	\thispageonlythisheader{\hfill DOE/NV/03624--0746}

	\begin{abstract}
		The models and weights of prior trained Convolutional Neural Networks (CNN) created to perform automated isotopic classification of time-sequenced gamma-ray spectra, were utilized to provide source domain knowledge as training on new domains of potential interest \cite{Moore19spie}.  The previous results were achieved solely using modeled spectral data.  In this work we attempt to transfer the knowledge gained to the new, if similar, domain of solely measured data.  The ability to train on modeled data and predict on measured data will be crucial in any successful data-driven approach to this problem space.
	\end{abstract}
	
	
	
	\section{INTRODUCTION} \label{sec:intro}
	Earlier research was motivated by the lack of robust automated identification algorithms for gamma-ray spectroscopy, specifically when the spectral statistics are low, including potentially low signal$/$noise ratios \cite{Ford18,Moore19spie}.  That earlier work was focused on automated gamma-ray spectral identification and new data schema for identification using Convolutional Neural Networks (CNN). This paper is focused on the problematic lack of available data sets in target domains of interest.  Although some good representative data for isotopes of interest may be available in small numbers most radio isotopes are not encountered in situ with great enough frequency to provide the large and diverse data sets needed for typical machine learning.  Typically, when large radiological data sets are collected the variety of sources found is very limited, with a few medical and industrial sources dominate the set; and as in our prior research significant reliance on simulated data is necessary.  This situation means that the machines trained over these sets of data will not be able to recognize the majority of possible radio-isotopes without some modification.

Our approach is to transfer the knowledge gained from these most common isotope classes to other isotopes so that ultimately this can be extended to the broader range of gamma spectroscopy in general.  While other efforts have been made to use adaptive algorithms for gamma-spectral identification, none of these have proved generalizable to all the domains of interest \cite{Portnoy04, Ford18, Jones14}.  As a proof of concept the source domain for this transfer learning will utilize both modeled and measured data \cite{URSC, novarray}.

	\section{DATA} \label{sec:data}
\begin{figure}[ht]
	\begin{center}
		\begin{tabular}{c}
			\includegraphics[height=9.5cm]{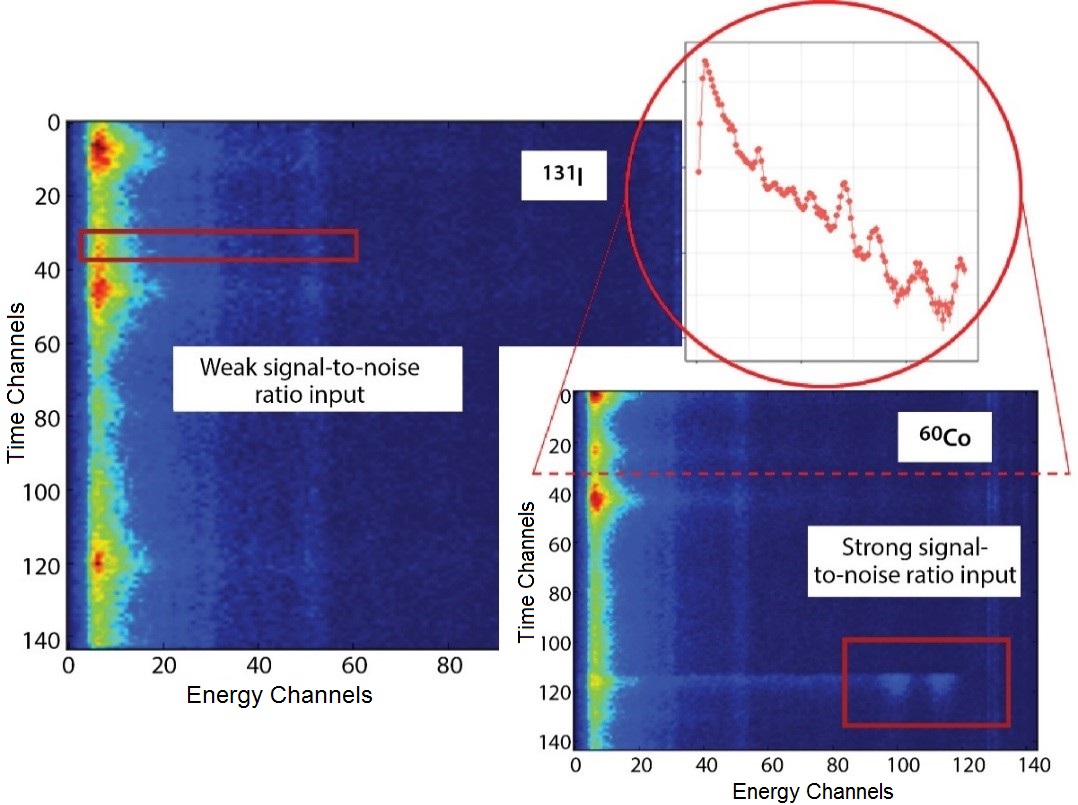}
		\end{tabular}		
	\end{center}
	\caption{Waterfall plots (false color of monochromatic images, in this case $144 \times  144$ pixels) illustrate the nature of the input used for training and testing the network. The signal-to-noise ratio may be strong or weak (the red boxes are meant to illustrate this, and do not correspond to an object detection box). A three-second time slice from the run (inset red circle) is meant to illustrate how the data are formatted \cite{Moore19spie}. }
	\label{fig:waterfall}
\end{figure}

	Typically gamma-ray data is collected in a `list-mode' structure with energy and arrival time of each photon recorded, but then stored in a compressed fashion. The stored data is usually in the form of energy-binned histograms with counts (photons) for a set integration time (e.g. - one second).  For low-resolution detectors (e.g - NaI) a typical binning is 1024 energy channels over 3 MeV range for 1 second; this binning schema is not ideal from an information content point-of-view, but still persists as the most frequently encountered in the community.  In this paper we are using 64~energy bins by 64~one~second time bins in the form of `waterfall' gray-scale images Fig.~\ref{fig:waterfall} \cite{CNNSDRD}.

\subsection{Simulated Data} \label{model}
The modeled data used came from a data competition (Urban Radiological Search Data Competition) run by Lawrence Berkeley National Laboratory (LBNL) for the DOE Office of Defense Nuclear Nonproliferation Research and Development \cite{URSC}. The portion of the modeled data used was the approximately $7000$ runs which were of long enough duration to meet the data structure requirements.  The modeled data is a simulated $2x4x16~inch$ \emph{NaI} detector moving through a simulated urban setting with changing backgrounds and periodic point sources of six different classes; background without a source present made up a seventh class for training purposes.  A more complete description of this data set exists in previous publications~\cite{URSC,Moore19spie}.

\subsection{Measured Data} \label{measurement}
The measured data used came from data collected in the stationary array in Virginia \cite{novarray}, and is referred to as the NOVArray data set. It is important to take note of the fact that unlike the simulated data the measured data is characterized by moving sources and fixed detectors, so that the backgrounds are continually changing in the initial training set. The data was dominated by medical isotopes with a few industrial sources mixed in and as discussed below some special nuclear material was available in the form of \emph{UF6}.  This data is also described in greater detail elsewhere, and can be made available to other researchers through the USDOE Office of Defense Nuclear Nonproliferation Research and Development (NA-22)~\cite{novarray,Moore19spie}.

\subsubsection{Confidence Level} \label{ssec: Nate}
Each \emph{hit} in the first thirty days of operation for the NOVArray dataset have been curated by a human spectroscopist; the analyst not only labeled the source encounters but also assigned a confidence for the label of \emph{low - medium - high}.  The analyst assigned a source label to all automatically detected anomalies. Frequently, spectra which generated a machine detection (e.g. not identification), that the human analyst would otherwise have considered background, were assigned to the most probable non-background class; this may explain why so many of the miss-assignments by the machines are into the background class.

	\section{ARCHITECTURES} \label{sec:arch}
	We used multiple machine learning algorithms, as discussed above, to classify the data set \cite{URSC}.  All of these involve the data set being pre-processed as described in the preceding two sections, although the exact image sizes varied by technique applied.  The architectures investigated were chosen because of their proven success with other image classification problems \cite{Lecun98, alex, resnet, inception}; much of the purely classification in this realm is dealt with in previous publications. Current work is focused on applying the methods developed to different radiological domains via transfer learning techniques, most importantly training on modeled data in order to predict on measured data.  Many different software libraries and machine learning frameworks were utilized in the building of these architectures \cite{tensorflow,keras,scik,caret}.

\subsection{Architectures Using Two Dimensional Data} \label{ssec:2D}

The two-dimensional data set (time \& energy) is shown in the waterfall image in Fig.~\ref{fig:waterfall}. Three convolutional architectures were used with the data in this form, and they are described in greater detail in previous work~\cite{Moore19spie}.  The first of these 2D convolutional architectures is based on early CNN architectures in the style of LeNet and VGG type \cite{Lecun98,alex, Simonyan15}, and this network is referred to throughout as \emph{Base CNN}.  Second is an architecture which uses `inception' modules \cite{inception}~and we refer to it simple as \emph{inception}.  And finally is a network that utilizes `residual connections' \cite{resnet,Sharma12}, which we refer to as \emph{residual}.

\subsection{Architectures Using One Dimensional Data} \label{ssec:1D}

A number of architectures were used to leverage the data in the energy domain only, as shown in the insert image in Fig.~\ref{fig:waterfall}. These methods use as input one second energy binned histograms.  This is the more traditional way of looking at the data, and is how a human analyst would approach the problem.  Automation of the problem using this type of data has a long history \cite{Portnoy04, Jones14, Hague19, Ford18}.  The types of architectures used here included: support vector machines (SVM), multilayered preceptrons (MLP), random forests (RF).

	\section{RESULTS} \label{sec:results}
	All results reflect multi-class accuracy, and all the metrics are the macro statistics across all relevant classes; care should be taken in interpreting the results with a full understanding of how things like class imbalance typically affect such statistics~\cite{metrics}.  A conscious decision was made to focus on the relevant 'basic' metrics (accuracy, precision, recall, specificity) which come directly from the 'raw' statistics (true-positives, false-positives, true-negatives, false-negatives); an exception to this appears in Table~\ref{table:transfer} were a couple composite statistics are shown.  Composite statistics can easily be constructed from the basic metrics for any reader who is interested. The relevant classes vary due to available data of a particular type - for example: the comparison of modeled to measured data only utilizes class common to both sets except where explicitly stated otherwise.  The transfer learning result is trained to one set (modeled) and then transferred to a different domain (measured).

\subsection{Training on Modeled Data and Testing on Measured Data} \label{ssec: modeled}
The URSC modeled data set was used to train the machines discussed in this section.  The machines where then tested on measured data from the NOVArray (Table~\ref{table:trainModel_testMeas}).  Only classes that overlapped both data sets where considered, and a small number of measured events where added to each training class - this was done to give the machines the opportunity to learn a new background.  Approximately 500 training events were available for each class of source with 20 measured events added to training - which were then excluded from the testing set.
\begin{table}[ht]
	\caption{Metrics for architectures trained on modeled data and tested on measured data.} 
	\centering 
	\begin{tabular}{l c c c c} 
		\hline 
		\rule{0pt}{3ex} \vspace{0.1cm} \textbf{Model} & \textbf{Accuracy\textsubscript{A}, \%} & \textbf{Precision\textsubscript{M}, \%} & \textbf{Recall\textsubscript{M}, \%} & \textbf{Specificity\textsubscript{M}, \%} \\ 
		\hline 
		\rule{0pt}{3ex} Base CNN & 94.26 & 71.49 & 68.70 & 95.99 \\
		\rule{0pt}{3ex} Inception & 97.60 & 87.25 & 95.38 & 97.62 \\
		\hline 
	\end{tabular}
	\label{table:trainModel_testMeas} 
\end{table}


\subsection{Trained and Tested on Measured Data} \label{ssec:all_measured}
For this section all the data considered is measured data for both training and testing; the training and testing was on a different subset of measured data - specifically medium confidence for training, and high \& low confidence for testing, in the case of Table~\ref{table:total_meas_only}. The confidence was assigned by the human analyst, as discussed in Sec.~\ref{ssec: Nate}.  The ensembling method is discussed in Sec.~\ref{ssec: temperature}.

The images used for the 2D machine results shown in this paper are $\mathit{64 x 64}$ pixel in the \emph{time~X~energy} domain, with a general binning scheme of one second time pixels and approximately 43~KeV equal energy binning from 40~KeV to 2800~KeV.  The binning schema was driven by overlapping restrictions on the two data sets used, and is clearly far from optimal.  A quick comparison of the results for this binning scheme and one at 4~Hz with 56 varying energy bins indicates a several percentage drop in accuracy for the binning scheme used.  The effect of this \emph{sparse energy information} content can not be fully appreciated in Table~\ref{table:total_meas_only}.  The 1D machines were binned at 256~energy bins over the same range as the 2D machines.
\begin{table}[ht]
	\caption{Metrics for architectures trained (medium confidence) and tested (high \& low confidence) measured data Sec.~\ref{ssec: Nate}. All metrics quoted are macro (across all classes), and displayed in \%.} 
	\centering 
	\begin{tabular}{l c c c c} 
		\hline 
		\rule{0pt}{3ex} \vspace{0.1cm} \textbf{Model} & \textbf{Accuracy\textsubscript{A}} & \textbf{Precision\textsubscript{M}} & \textbf{Recall\textsubscript{M}} & \textbf{Specificity\textsubscript{M}} \\ 
		\hline \\ 
		\multicolumn{5}{l}{\emph{Two Dimensional Methods: (sparse energy information content - 64 energy bins)}}\\
		\rule{0pt}{3ex} Base CNN & 95.23 & 74.47 & 70.98 &   96.57 \\
		\rule{0pt}{3ex} Inception & 97.14 & 90.32 & 81.43 & 97.66 \\
		\rule{0pt}{3ex} Residual Connections & 96.26 & 85.54 & 75.20 & 97.04 \\
		\rule{0pt}{3ex} Ensemble of 2D methods (Sec.~\ref{ssec: temperature}) & 97.28 & 90.45 & 81.90 & 97.79 \\
		\\
		\multicolumn{5}{l}{\emph{One Dimensional Methods: (dense energy information content- 256 energy bins)}}\\
		\rule{0pt}{3ex} Support Vector Machine (SVM) & 97.51 & 93.07 & 81.85 & 97.93 \\
		\rule{0pt}{3ex} Multi-Layered Preceptron (MLP) & 97.22 & 88.67 & 83.02 & 97.88 \\
		\rule{0pt}{3ex} Ensemble of 1D w/o temperature & 97.54 & 91.48 & 82.90 & 98.03 \\
		\rule{0pt}{3ex} Ensemble of 1D with temperature & 97.11 & 92.17 & 78.91 & 97.55 \\
		\\
		\multicolumn{5}{l}{\emph{One \& Two Dimensional Methods Combined}}\\
		\rule{0pt}{3ex} Full Ensemble & 97.67 & 92.07 & 83.33 & 98.14 \\

		\hline 
	\end{tabular}
	\label{table:total_meas_only} 
	
\end{table}

\subsection{Incorporation of Different Class and Location} \label{ssec: uranium}
In addition to measurement data being collected around northern Virginia some data was collected outside a uranium processing facility - a completely different background region.  This data set consists of mostly low or un-enriched uranium hexafloride, some medical and industrial sources where seen during this other data collect; this data was combined with the NOVArray data set and a new uranium class added into the training.  Results of training over the combined collections using the 2D machine learning methods is shown in Table~\ref{table:frama}.  Low precision on the uranium class is due to a great extent to class imbalance; although less than one-percent of Tc99m was miss-classified as U238, however this was a significant number relative to the number of U238 testing samples.  It should be noted that what we are calling \emph{miss-classifications} assume that the human analyst made no errors.  It must also be noted when assessing the statistics in Table~\ref{table:frama} that the small number of events in the U238 class mean that slight variations have significant effect on the statistics. Unlike the case in Sec.~\ref{ssec:all_measured} above in this case the train/test split is 75/25 percent randomly drawn without regard to the confidence level assignment by the human spectroscopist.

\begin{table}[ht]
	\caption{Metrics for architectures trained and tested measured data Sec.~\ref{ssec: Nate}. All metrics quoted are macro (across all classes), and displayed in \%.} 
	\centering 
	\begin{tabular}{l c c c c} 
		\hline 
		\rule{0pt}{3ex} \vspace{0.1cm} \textbf{Model} & \textbf{Accuracy\textsubscript{A}} & \textbf{Precision\textsubscript{M}} & \textbf{Recall\textsubscript{M}} & \textbf{Specificity\textsubscript{M}} \\ 
		\hline \\ 
		\multicolumn{5}{c}{\emph{Results across full set of classes}}\\
		\rule{0pt}{3ex} Base CNN & 95.54 & 78.15 & 77.03 &   97.58 \\
		\rule{0pt}{3ex} Inception & 97.41 & 84.09 & 86.05 & 98.49 \\
		\rule{0pt}{3ex} Residual Connections & 96.21 & 81.11 & 71.68 & 97.90 \\
		\rule{0pt}{3ex} Best ensemble & 97.31 & 84.46 & 86.33 & 98.45 \\
		\\
		\multicolumn{5}{c}{\emph{Results specific to Uranium class only}}\\
		\rule{0pt}{3ex} Best ensemble & 98.96 &	70.13 &	96.43 &	99.016 \\

		\hline 
	\end{tabular}
	\label{table:frama} 
	
\end{table}
\subsection{Training on Modeled Data and Transfer Knowledge to Measured Domain} \label{ssec: transfer}
There are a number of ways to approach the problem of transferring the knowledge gained from one domain into another domain.  The goal of the transfer in our case is to gain predictive power for classes of radioactive source spectra that are difficult or impossible to acquire in situ spectra in significant numbers, but are still of primary concern from the perspective of detection in operationally realistic environments.

We found that one of the simplest ways to achieve this was to simply \emph{seed} a class of little interest with a small number of examples from the source of interest.  The problem with this approach is that you are now open to significant false positives from a source class that is of no concern.  Ideally in the case of simulated data you could just use the measured examples to augment the modeled data set.

\begin{table}[ht]
	\caption{Architectures trained on modeled data first and the weights are used to initialize the training on measured data Sec.~\ref{ssec: Nate}. In this instance composite metrics are shown explicitly.  All metrics quoted are macro (across all classes), and displayed in \%.}
	\centering
	\begin{tabular}{l c c c c c c}
		\hline		
		\rule{0pt}{3ex} \vspace{0.1cm} \textbf{Method} & \textbf{Accuracy\textsubscript{A}} & \textbf{Precision\textsubscript{M}} & \textbf{Recall\textsubscript{M}} & \textbf{Specificity\textsubscript{M}} & \textbf{f1 Score} & \textbf{J Statistic} \\
		\hline \\
		\multicolumn{7}{c}{\emph{All Class Comparison  - Inception Architecture with \& without Transfer Learning}}\\
		\rule{0pt}{3ex} without Transfer & 96.75 & 78.50 & 80.48 & 98.09 &  79.48 & 78.57 \\
		\rule{0pt}{3ex} with Transfer & 96.72 & 76.94 & 82.20 &  98.03 & 79.48 & 80.24 \\
		\\
		\multicolumn{7}{c}{\emph{Uranium Class Only - Inception Architecture with \& without Transfer Learning}}\\
		\rule{0pt}{3ex} without Transfer & 98.49 & 67.52 & 68.13 & 99.22 & 67.82 & 67.35 \\
		\rule{0pt}{3ex} with Transfer & 98.50 & 62.79 & 88.13 &	98.75 &	73.33 & 86.88 \\				
		\hline
	\end{tabular}
	\label{table:transfer}	
\end{table}

In addition to these approaches we also trained entirely in one source domain (modeled data) and transferred an almost entirely frozen architecture to a different target domain (measured data), only re-training the classification layer or the last couple fully connected layers.  This proved to be the most difficult and least robust approach.  The best results were achieved by training on the modeled data and the trained model \& weights are used to initialize the training on the new domain (Table~\ref{table:transfer})

Although it is unclear how much benefit may be gained by this type of transfer learning the best result for the U238 class specifically was obtained in this way.  Further investigation is warranted as more data becomes available.

\subsection{Calibration \& Ensembling} \label{ssec: temperature}
Deep learning models typically output some form of softmax vector which may be interpreted as a probability vector where the individual logits into the softmax approximately represent the confidence level of the input distribution being associated with the given class.  The interpretation of softmax as a true probability vector generally becomes less certain as the depth and complexity of the model increases.  Ways of calibrating a given model to a true probability vector is an area of active research \cite{reliability}.  For the purposes of this research we only investigate the effect of softening the softmax distribution via the inclusion of temperature ($T$) into the softmax.
\[ \sigma (z_i)^{(k)} = \frac{exp ({z_i^{(k)}/T}) }{\sum_{j=0}^{K} exp ({z_i^{(j)}/T}) }, \hspace{0.75 cm} for \hspace{0.25 cm} j=1,\mathellipsis ,K ,\] \\
where the ${k^{th}}$ element of the softmax ($\sigma^{k}$) is \emph{softened} by the temperature ($T$). Beyond the argument for making the softmax outputs more consistent with true confidence intervals this method can also yield improved knowledge gain from the relationships of the "non-winning" (non-maximum) softmax output.

\begin{figure}[ht]
	\begin{center}
		\begin{tabular}{c}
			\includegraphics[height=9.5cm]{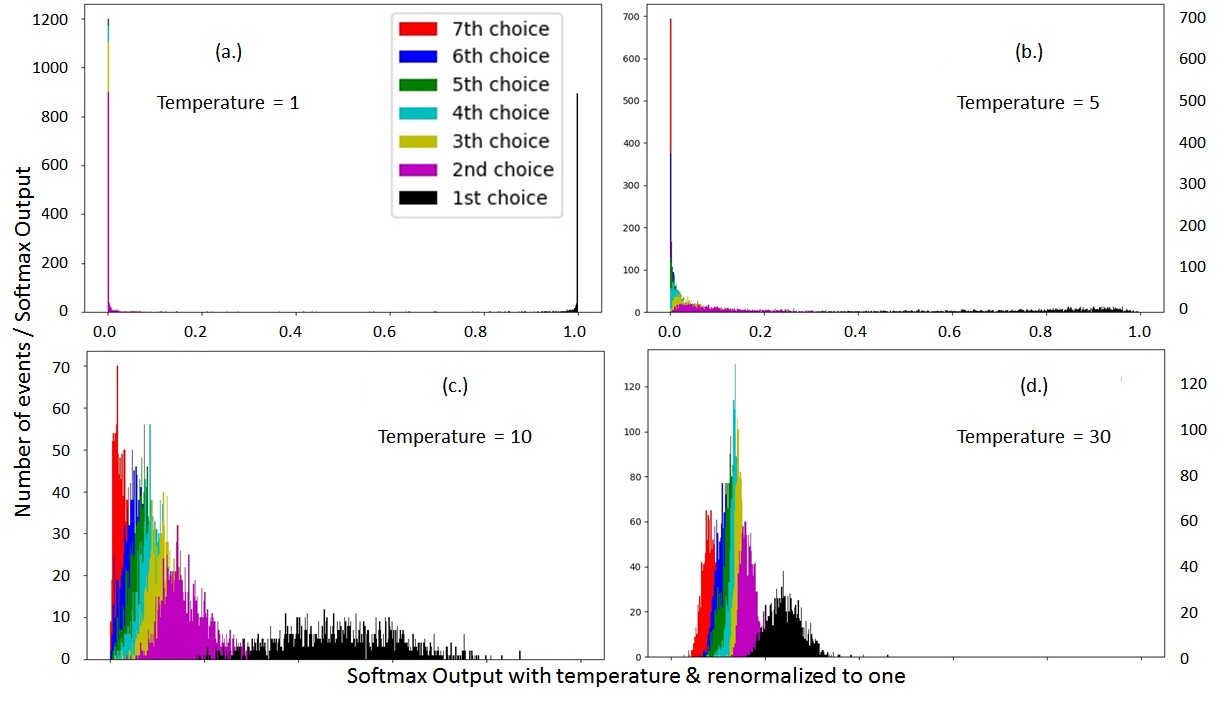}
		\end{tabular}		
	\end{center}
	\caption{The temperature softened softmax output of a deep CNN is shown for a variety of temperatures for the same sample of testing data.  The setting of temperature equal to one implies no softening. }
	\label{fig:temp}
\end{figure}

The other elements of the softmax distribution may yield information about the class distribution in the form of `dark knowledge' \cite{Hinton15}.  This dark knowledge might otherwise be lost if the softmax distribution is not softened as in the upper-left corner of Fig.~\ref{fig:temp}.  Care must be taken if one wishes to preserve the concept of a probability distribution in that scaling by a constant in the exponential distribution will clearly mean that the elements will no longer sum to one unless specifically re-normalized; as has been done here, so that the softmax distribution of any given output ($y$) still sums to one.

This technique was used throughout the work to ensemble the different model outputs for all the results shown.  More work needs to be done in tuning the temperature parameter.  The method used to select the temperature parameter's value was somewhat ad hoc, the distribution of softmax outputs for each output vector $y_i$, where $i$ is over the number of classes, is histogrammed at various values of temperatures Fig.~\ref{fig:temp}.  A best guess of an appropriate temperature hyper-parameter is made with the intent of gaining as much dark knowledge as possible; unfortunately a more rigorous selection would have required a larger curated data set.

	\section{CONCLUSION} \label{sec:conclusion}
	Although other work has demonstrated the utility of machine learning techniques in gamma ray spectroscopy \cite{Portnoy04,Sharma12,Jones14,Ford18,Moore19spie}, these methods are just starting to take advantage of the expanding availability of large data sets; where some of our prior work laid the ground work for methods shown here few researches have been able to make use of large curated measured spectral data sets in the gamma domain to validate their results.  

The techniques presented demonstrate that simulated data can be successfully combined with measured data to create widely applicable and robust methods; this should be leveraged to engineer more robust fielded algorithms for automated identification.  Modeled source spectra should be injected onto measured backgrounds for this purpose.

As far as the authors are aware the application of techniques for extracting additional dark knowledge~\cite{Hinton15} from the softmax have not been applied in this domain, and provides a potentially valuable tool for ensembling to improve performance~\cite{Moore20}.

	\acknowledgements
	We would like to acknowledge the Department of Energy’s support through the Office of Defense Nuclear Nonproliferation Research and Development NA-22 and the contributors to the \cite{URSC} (`Urban Radiological Search Data Competition') for the data set.  And we would also like to acknowledge the support of the Site Directed Research and Development (SDRD) program for the National Nuclear Security Site (NNSS).  We would also like to acknowledge the extensive work done in the fielding and collection of the data, particularly by: Andr\'e Butler, Emily Jackson, and Ron Wolff.

This manuscript has been authored by Mission Support and Test Services, LLC, under Contract No. DE-NA0003624 with the U.S. Department of Energy and supported by the Site-Directed Research and Development Program, National Nuclear Security Administration, USDOE Office of Defense Nuclear Nonproliferation Research and Development (NA-22). The United States Government retains and the publisher, by accepting the article for publication, acknowledges that the United States Government retains a non-exclusive, paid-up, irrevocable, worldwide license to publish or reproduce the published form of this manuscript, or allow others to do so, for United States Government purposes. The U.S. Department of Energy will provide public access to these results of federally sponsored research in accordance with the DOE Public Access Plan (http://energy.gov/downloads/doe-public-access-plan). The views expressed in the article do not necessarily represent the views of the U.S. Department of Energy or the United States Government. DOE/NV/03624--0746.

	\bibliography{references}   

\begin{thebibliography}{10}
\providecommand{\url}[1]{#1}
\csname url@samestyle\endcsname
\providecommand{\newblock}{\relax}
\providecommand{\bibinfo}[2]{#2}
\providecommand{\BIBentrySTDinterwordspacing}{\spaceskip=0pt\relax}
\providecommand{\BIBentryALTinterwordstretchfactor}{4}
\providecommand{\BIBentryALTinterwordspacing}{\spaceskip=\fontdimen2\font plus
\BIBentryALTinterwordstretchfactor\fontdimen3\font minus
  \fontdimen4\font\relax}
\providecommand{\BIBforeignlanguage}[2]{{%
\expandafter\ifx\csname l@#1\endcsname\relax
\typeout{** WARNING: IEEEtran.bst: No hyphenation pattern has been}%
\typeout{** loaded for the language `#1'. Using the pattern for}%
\typeout{** the default language instead.}%
\else
\language=\csname l@#1\endcsname
\fi
#2}}
\providecommand{\BIBdecl}{\relax}
\BIBdecl

\bibitem{Moore19spie}
\BIBentryALTinterwordspacing
E.~T. Moore, W.~P. Ford, E.~J. Hague, and J.~L. Turk, ``{An Application of CNNs
  to Time Sequenced OneDimensional Data in Radiation Detection},'' ser.
  Proceedings of the SPIE, vol. 10986, 2019. [Online]. Available:
  \url{https://arxiv.org/abs/1908.10887}
\BIBentrySTDinterwordspacing

\bibitem{Ford18}
W.~P. Ford, E.~Hague, T.~McCullough, E.~T. Moore, and J.~Turk, ``Threat
  determination for radiation detection from the remote sensing laboratory,''
  ser. Proceedings of the SPIE, vol. 10644, 2018, p. 106440G.

\bibitem{Portnoy04}
D.~Portnoy, P.~Bock, P.~C. Heimberg, and E.~T. Moore, ``Using alisa for
  high-speed classification of the components and their concentrations in
  mixtures of radioisotopes,'' ser. Proceedings of the SPIE, vol. 5541, 2004,
  pp. 1--10.

\bibitem{Jones14}
A.~Jones, ``Machine learning for classification and visualisation of
  radioactive substances for nuclear forensics,'' \emph{Proceedings of the
  Information Analysis Technologies, Techniques and Methods for Safeguards,
  Nonproliferation and Arms Control Verification Workshop}, May 2014.

\bibitem{URSC}
\BIBentryALTinterwordspacing
DOE/NNSA, ``Urban radiological search data competition,'' ({Accessed:~2018}).
  [Online]. Available: \url{https://datacompetitions.lbl.gov/competition/1/}
\BIBentrySTDinterwordspacing

\bibitem{novarray}
{DOE/NNSA - Remote Sensing Laboratory}, ``Northern {V}irginia {A}rray
  ({NoVArray}),'' { {\it came online:~{D}ecember 2018.} Contact authors for
  information.}

\bibitem{CNNSDRD}
{E. T. Moore, W. P. Ford, E. J. Hague, and J. L. Turk}, ``Algorithm development
  for targeted isotopics,'' \emph{Site-Directed Research and Development - FY
  2018}, 2019, {NNSS}, Las Vegas, Nevada.

\bibitem{Lecun98}
Y.~LeCun, L.~Bottou, Y.~Bengio, and P.~Haffner, ``Gradient-based learning
  applied to document recognition,'' ser. IEEE Proc., vol. 86, 11, 1998, p.
  2278–2324.

\bibitem{alex}
A.~Krizhevsky, I.~Sutskever, and G.~E. Hinton, ``Imagenet classification with
  deep convolutional neural networks,'' ser. Proc. NIPS, vol. (CVPR), 2012, p.
  1106–1114.

\bibitem{resnet}
K.~He, X.~Zhang, S.~Ren, and J.~Sun, ``Deep residual learning for image
  recognition,'' ser. IEEE Conference on Computer Vision and Pattern
  Recognition (CVPR), June 2016, pp. Las Vegas, Nevada.

\bibitem{inception}
C.~Szegedy, W.~Liu, Y.~Jia, P.~Sermanet, S.~Reed, D.~Anguelov, D.~Erhan,
  V.~Vanhoucke, and A.~Rabinovich, ``Going deeper with convolutions,'' ser.
  IEEE Conference on Computer Vision and Pattern Recognition (CVPR), June 2015,
  pp. 1--9.

\bibitem{tensorflow}
\BIBentryALTinterwordspacing
M.~Abadi \emph{et~al.}, ``{TensorFlow}: Large-scale machine learning on
  heterogeneous systems,'' ({Accessed:~2016}). [Online]. Available:
  \url{https://www.tensorflow.org/}
\BIBentrySTDinterwordspacing

\bibitem{keras}
F.~Chollet \emph{et~al.}, ``Keras,'' url\{https://github.com/fchollet/keras\},
  2015.

\bibitem{scik}
F.~Pedregosa \emph{et~al.}, ``Scikit-learn: Machine learning in python,''
  \emph{Journal of Machine Learning Research (JMLR)}, vol.~12, pp. 2825--2830,
  2011.

\bibitem{caret}
M.~Kuhn, ``Building predictive models in r using the caret package,'' \emph{J.
  Statistical Software}, vol.~28, p.~5, 2008.

\bibitem{Simonyan15}
K.~Simonyan and A.~Zisserman, ``Very deep convolutional networks for
  large-scale image recognition,'' 2015.

\bibitem{Sharma12}
S.~Sharma, C.~Bellinger, N.~Japkowicz, R.~Berg, and K.~Ungar, ``Anomaly
  detection in gamma ray spectra: A machine learning perspective,'' \emph{2012
  IEEE Symposium on Computational Intelligence for Security and Defence
  Applications}, July 2012, iEEE Xplore: 31 August 2012.

\bibitem{Hague19}
\BIBentryALTinterwordspacing
E.~J. Hague, W.~P. Ford, E.~T. Moore, and J.~L. Turk, ``{A comparison of
  adaptive and template matching techniques for radio-isotope
  identification},'' ser. Proceedings of the SPIE, vol. 10986, 2019. [Online].
  Available: \url{https://doi.org/10.1117/12.2519062}
\BIBentrySTDinterwordspacing

\bibitem{metrics}
\BIBentryALTinterwordspacing
M.~Sokolova and G.~Lapalmes, ``A systematic analysis of performance measures
  for classification tasks,'' \emph{Science Direct: Information Processing \&
  Management}, vol. 45 issue 4, pp. 427--437, 2009. [Online]. Available:
  \url{https://doi.org/10.1016/j.ipm.2009.03.002}
\BIBentrySTDinterwordspacing

\bibitem{reliability}
C.~Guo, G.~Pleiss, Y.~Sun, and K.~Q. Weinberger, ``On calibration of modern
  neural networks,'' ser. Proceedings of the 34th ICML, vol.~70, 2017, pp.
  1321--1330.

\bibitem{Hinton15}
G.~Hinton, O.~Vinyals, and J.~Dean, ``Distilling the knowledge in a neural
  network,'' \emph{{https://arxiv.org/abs/1503.02531}}, 2015, {\it Submitted on
  9 Mar 2015}.

\bibitem{Moore20}
E.~Moore, ``Utilizing dark knowledge to ensemble outputs for spectral
  identification,'' \emph{arXiv}, 2020, {\emph{in Draft}}.

\end{thebibliography}
	\bibliographystyle{ieeetran}   
	
	
\end{document}